\documentclass[aps,prl,reprint,superscriptaddress,showpacs]{revtex4-1}
\usepackage{graphicx}
\usepackage{epsfig}
\usepackage{epstopdf}
\usepackage{amsmath}
\usepackage{color}
\usepackage[colorlinks=true, citecolor=blue, linkcolor=blue, urlcolor=blue]{hyperref}
\begin{document}
\title{Adsorbate Electric Fields on a Cryogenic Atom Chip}
\author{K.S.~Chan}
\affiliation{Division of Physics and Applied Physics, Nanyang Technological 
University,  21 Nanyang Link, Singapore 637371, Singapore}
\author{M.~Siercke}
\affiliation{Division of Physics and Applied Physics, Nanyang Technological 
University,  21 Nanyang Link, Singapore 637371, Singapore}
\affiliation{Centre for Quantum Technologies, National University of 
Singapore, 3 Science Drive 2, Singapore 117543, Singapore}
\author{C.~Hufnagel}
\affiliation{Division of Physics and Applied Physics, Nanyang Technological 
University,  21 Nanyang Link, Singapore 637371, Singapore}
\affiliation{Centre for Quantum Technologies, National University of 
Singapore, 3 Science Drive 2, Singapore 117543, Singapore}
\author{R.~Dumke}
\email{rdumke@ntu.edu.sg}
\affiliation{Division of Physics and Applied Physics, Nanyang Technological 
University,  21 Nanyang Link, Singapore 637371, Singapore}
\affiliation{Centre for Quantum Technologies, National University of 
Singapore, 3 Science Drive 2, Singapore 117543, Singapore}
\date{\today}
\begin{abstract}
We investigate the behaviour of electric fields originating from adsorbates deposited on a cryogenic atom chip as it is cooled from room temperature to cryogenic temperature. Using Rydberg electromagnetically induced transparency we measure the field strength versus distance from a $1 \; \mathrm{mm}$ square of YBCO patterned onto a YSZ chip substrate. We find a localized and stable dipole field at room temperature and attribute it to a saturated layer of chemically adsorbed rubidium atoms on the YBCO. As the chip is cooled towards 83 $\mathrm{K}$ we observe a change in sign of the electric field as well as a transition from a localized to a delocalized dipole density. We relate these changes to the onset of physisorption on the chip surface when the van der Waals attraction overcomes the thermal desorption mechanisms. Our findings suggest that, through careful selection of substrate materials, it may be possible to reduce the electric fields caused by atomic adsorption on chips, opening up experiments to controlled Rydberg-surface coupling schemes.
\end{abstract}
\maketitle
In recent years, atom chips \cite{RevModPhys.79.235,AtomChips_Book} have developed into a major platform 
for the investigation of atomic quantum states. Offering precise control of neutral atoms combined with long coherence times \cite{PhysRevLett.92.203005} make them a promising candidate for quantum simulation applications. Furthermore, atom chips are scalable and can be integrated with other systems, such as solid-state devices, to form hybrid quantum systems \cite{PhysRevLett.92.063601,NaturePhysics.9.636}.   

To establish fast quantum gates for neutral atoms, it was proposed early on to use atomic Rydberg states \cite{PhysRevLett.85.2208,PhysRevLett.87.037901}. The strong and well controllable dipolar interaction of these states allows a fast coupling between atoms, making them an ideal choice for quantum processing protocols \cite{RevModPhys.82.2313}. Moreover, the large polarizability of Rydberg states provides a strong coupling to electric fields, which can be used to couple them to on-chip solid state devices 
\cite{PhysRevLett.100.170501,PhysRevA.79.040304,PhysRevLett.108.063004}. However, due to this strong interaction, Rydberg states are also very prone to transition shifts caused by residual electric fields. 
It has been shown, that close to chip structures a major cause of these electrostatic fields is given by surface adsorbed atoms \cite{cornell_2004,cornell_2007,spreeuw_2010,cs_adams_ads,fortagh_2012}. 

\begin{figure}[h]
\centering
\includegraphics[width=.48\textwidth]{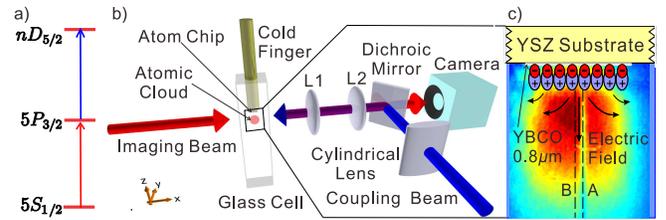}
\caption{\label{fig:exp_setup}(color online). a) Relevant energy level structure for $^{87}$Rb. b) Schematic diagram of the experimental setup. The probe beam (red) couples the ground state $5S_{1/2}$ to the intermediate state $5P_{3/2}$ while the coupling beam (blue) couples the intermediate state to an $nD_{5/2}$ Rydberg state. To probe the electric field close to the chip surface, the coupling beam is focused to a vertical line. c) Absorption imaging of an atomic cloud below the cryogenic chip (1 mm YBCO square on a 20 mm YSZ substrate). The central line of low optical density is caused by EIT. With spatial changes in the transparency signal the electric field caused by the adsorbates can be mapped.}
\end{figure}

In general, there are two types of adsorption processes, categorized by their bonding type. Chemisorption involves the transfer of charge from adatoms to the substrate or vice versa depending on the relative value of the work function of the substrate and the ionization energy of the adatom \cite{book_nanophysics_clusters}. Alternatively, atoms can be adsorbed to the substrate through the mediation of van der Waals forces i.e. a bond formed from the dipole-dipole interaction between the adatom and the resultant induced image dipole inside the substrate \cite{book_langmuir_isobar}. Due to the much weaker binding strength of the van der Waals bond compared to a chemical bond, this physisorption has not been significant in previous works, since all studies have been carried out at room temperature \cite{cornell_2004,cornell_2007,spreeuw_2010,fortagh_2012}. However, for hybrid quantum systems where atoms are coupled to superconducting devices, the atom chip will be cooled to cryogenic temperatures where such van der Waals bonding can no longer be neglected. 

In this paper, we investigate the temperature dependence of the atom-surface bonding on a cryogenic atom chip. Using electromagnetically induced transparency (EIT) \cite{RevModPhys.77.633} in  $^{87}$Rb for Rydberg spectroscopy, we determine the electric field produced by adatoms in a temperature range of 83 K to room temperature. We compare our findings to the theoretical temperature dependence of physisorption and chemisorption to deduce the relative strength of the two mechanisms in the different temperature regimes and to investigate ways by which electric fields arising from adsorbates may be minimized. 

We model a layer of adsorbates by a square dipole layer with side length $d$ on top of the substrate. The simulated electric field perpendicular to the centre of the dipole layer is obtained from the summation of the electric field from two opposite but constant surface charge square sheets separated by $1 \; \mathring{\text{A}}$, much less than the distance between the surface and the electric field probe (a gas of ultracold neutral atoms):
\begin{equation}
E(z)=\frac{2\sigma}{\pi\epsilon_0} \left[\tan^{-1}\sqrt{1+\frac{d^2}{2z^2}}-\tan^{-1}\sqrt{1+\frac{d^2}{2(z+1\mathring{\text{A}})^2}}\right],
\label{equa_dipole}
\end{equation}
where $\epsilon_0$ is the electric permittivity of free space.
In this limit the exact choice of $1 \; \mathring{\text{A}}$ will not change the value of the dipole density. The side length $(d)$ and the surface charge density $(\sigma)$ are experimentally determined parameters.
 
Our experimental setup is illustrated in Fig. \ref{fig:exp_setup}b: A cloud of ultracold $^{87}\mathrm{Rb}$ is prepared in proximity of a $1 \; \mathrm{mm}$ YBCO square patterned onto a 20 mm YSZ square substrate. Details of the preparation of the ultracold gas is discussed in \cite{atoms-vortices,programmable_vortices,self-sufficient-exp}. The cloud is illuminated by two counter propagating laser beams resonant with the $5S_{1/2}\rightarrow 5P_{3/2}$ (probe beam) and $5P_{3/2} \rightarrow nD_{5/2}$ (coupling beam) transitions respectively, where $n$ indicates the principal quantum number. The coupling beam is focused to a  $100 \; \mathrm{\mu m}\times2.5 \; \mathrm{mm}$ line perpendicular to the YBCO square structure. As shown in Fig. \ref{fig:exp_setup}a, when both beams are tuned to resonance, absorption of the probe beam is inhibited due to EIT \cite{RevModPhys.77.633}, resulting in a line of decreased optical density of the atomic cloud at the position of the coupling beam focus. Due to the large polarizability of the $n$D Rydberg state used, small electric fields split the $|m_j|$ sublevels allowing a measurement of the spatial dependence of the electric field near the surface by determining the EIT resonance frequencies vs. distance. We define the optical density ratio (ODR) by dividing column A (Fig. \ref{fig:exp_setup}a) inside by column B outside of the coupling beam region.

\begin{figure}[ht]
\includegraphics[width=.48\textwidth]{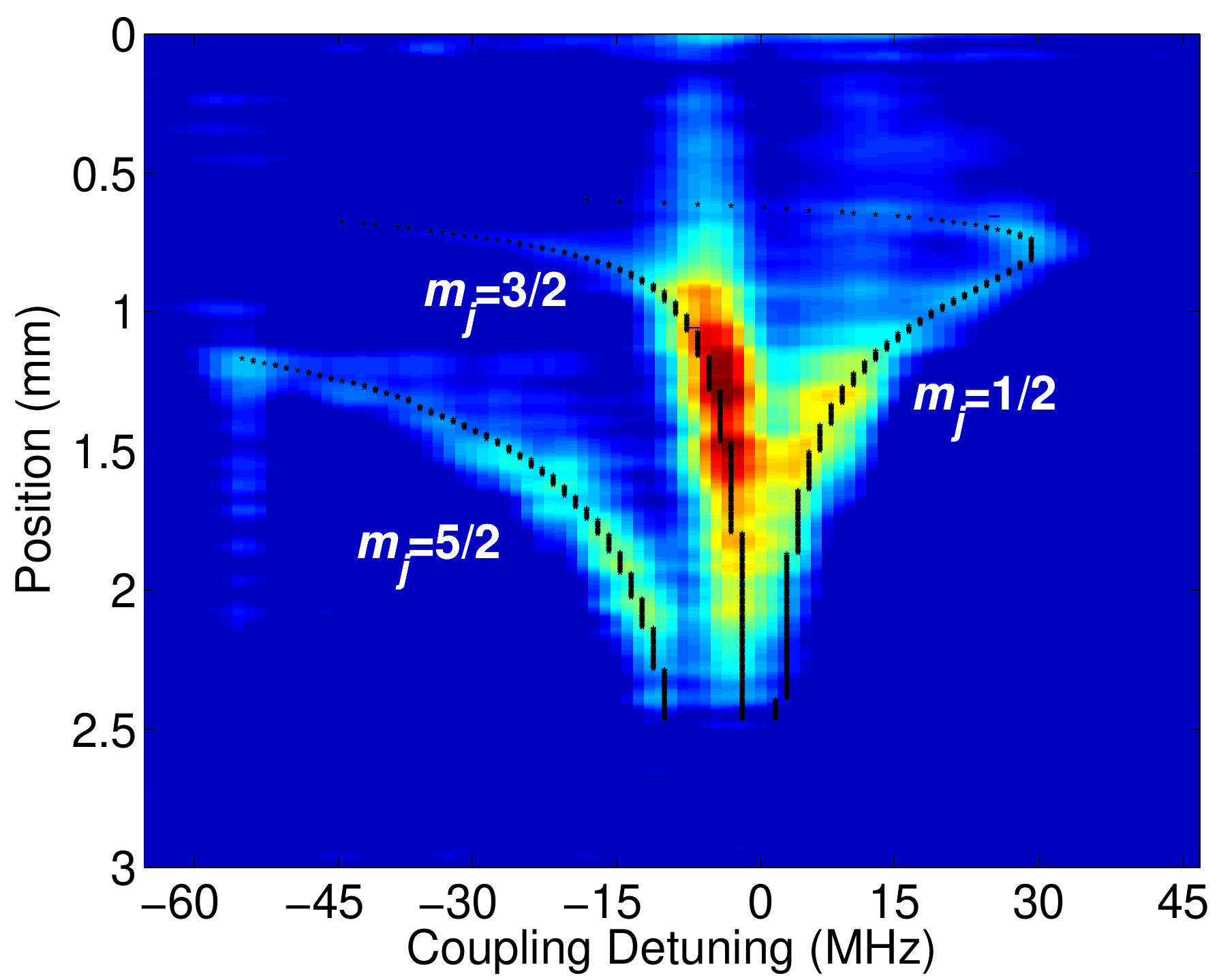}
\caption{\label{fig:room_temp_58D}(color online). Rydberg spectroscopy of the 58$\textit{D}_{5/2} $ $ |m_j|=5/2,3/2,1/2$ states obtained at room temperature. The horizontal axis is the detuning of the coupling laser from the $5P_{3/2}$ to 58$\textit{D}_{5/2}$ transition. The vertical axis is the distance to the substrate. The degeneracy of the  58$\textit{D}_{5/2}$ manifold is lifted by the presence of an electric field. A fit to the data using Eq.(\ref{equa_dipole}) (black line) gives a size of the adsorbed layer of $1\mathrm{mm}$ with a surface dipole density of $3.7\times 10^{-12} \; \text{C m$^{-1}$}$ and an offset electric field of $0.14  \; \text{V/cm}$.}
\end{figure}

Fig. \ref{fig:room_temp_58D} shows the ODR for atoms coupled to the 58$\textit{D}_{5/2}$ state at room temperature. The chip surface is indicated by a black line at the origin. We observe three lines that merge far away from the surface, corresponding to the $|m_j|=1/2$, $|m_j|=3/2$ and $|m_j|=5/2$ manifolds of the $58D_{5/2}$ state. The spatial behaviour of the line splitting in Fig. \ref{fig:room_temp_58D} suggests an electric field close to the chip surface that decreases over a length scale of $\approx 1 \; \mathrm{mm}$. The strong spatial dependence of the field over such a short distance implies that it is produced by a source of similar size. Fitting the spectrum using Eq.(\ref{equa_dipole}) and the Stark shift (given in Ref. \cite{zimmerman_stark}) for the $58D_{5/2}$ state gives a surface dipole density of $3.7 \; \times 10^{-12} \; \text{C m$^{-1}$}$ spread over a $1 \; \mathrm{mm}$ square. The data analysis indicates a small offset field of $0.14 \; \text{ V/cm}$, suggesting the presence of a constant background field.

The $1 \; \mathrm{mm}$ size of the dipole distribution extracted from the fit to Fig. \ref{fig:room_temp_58D} is caused by rubidium atoms chemisorbed onto the $1 \; \mathrm{mm}$ YBCO square. The strength of the chemisorbed dipole depends on the difference between the work function of the substrate and the ionization energy of the atom. The work function of YBCO (4.5 eV) \cite{ybco_work_func} differs from the rubidium ionization energy (4.2 eV), potentially giving rise to strong chemisorbed dipoles, while the work function of our YSZ substrate (4.34 eV) \cite{ysz_work_function} is closer to the ionization energy of a rubidium atom. Still, chemisorption onto the YSZ substrate may be responsible for the small offset field observed in Fig. \ref{fig:room_temp_58D}.

\begin{figure}[h]
\centering
\includegraphics[width=.48\textwidth]{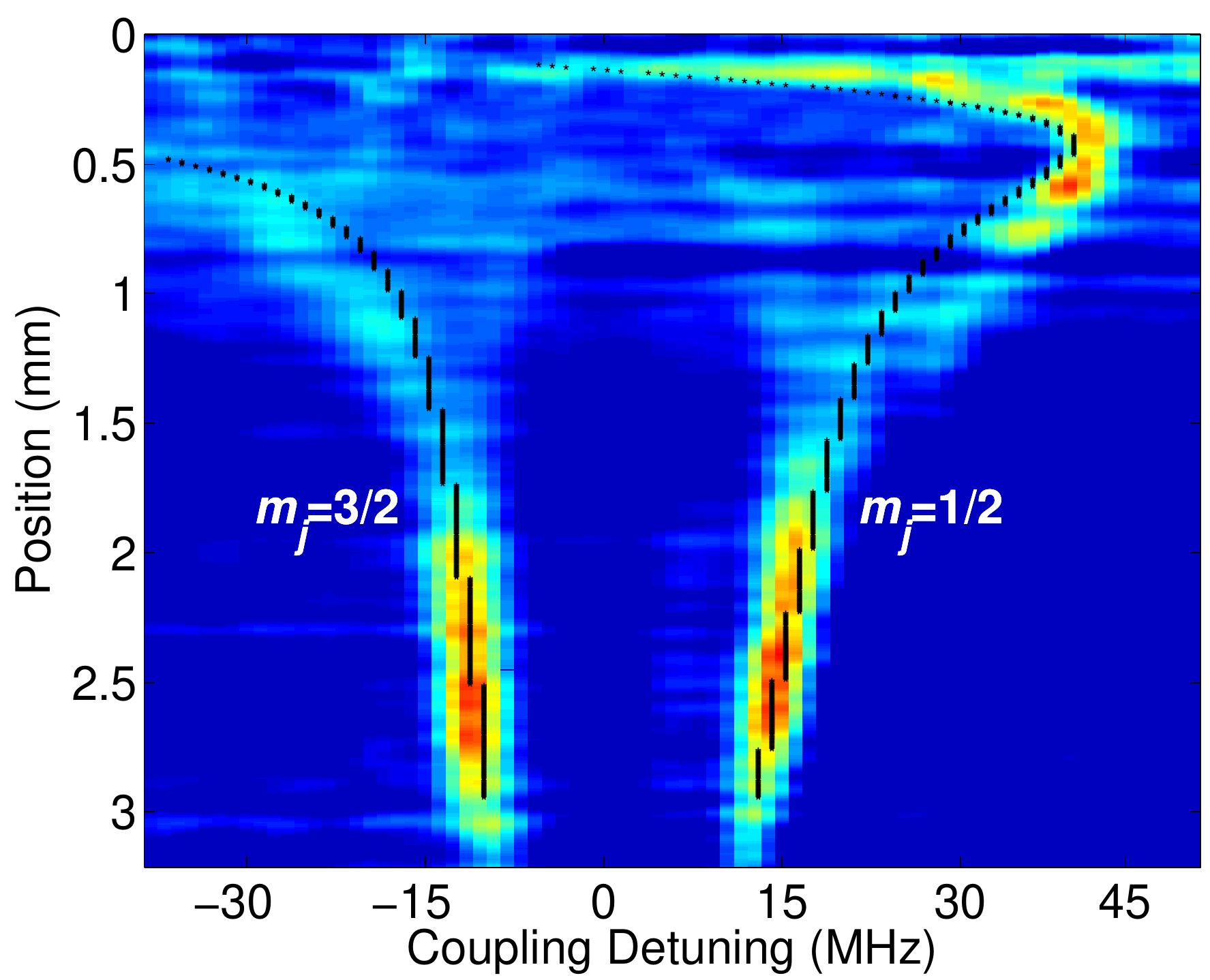}\llap{\raisebox{.95cm}{\includegraphics[scale=0.135]{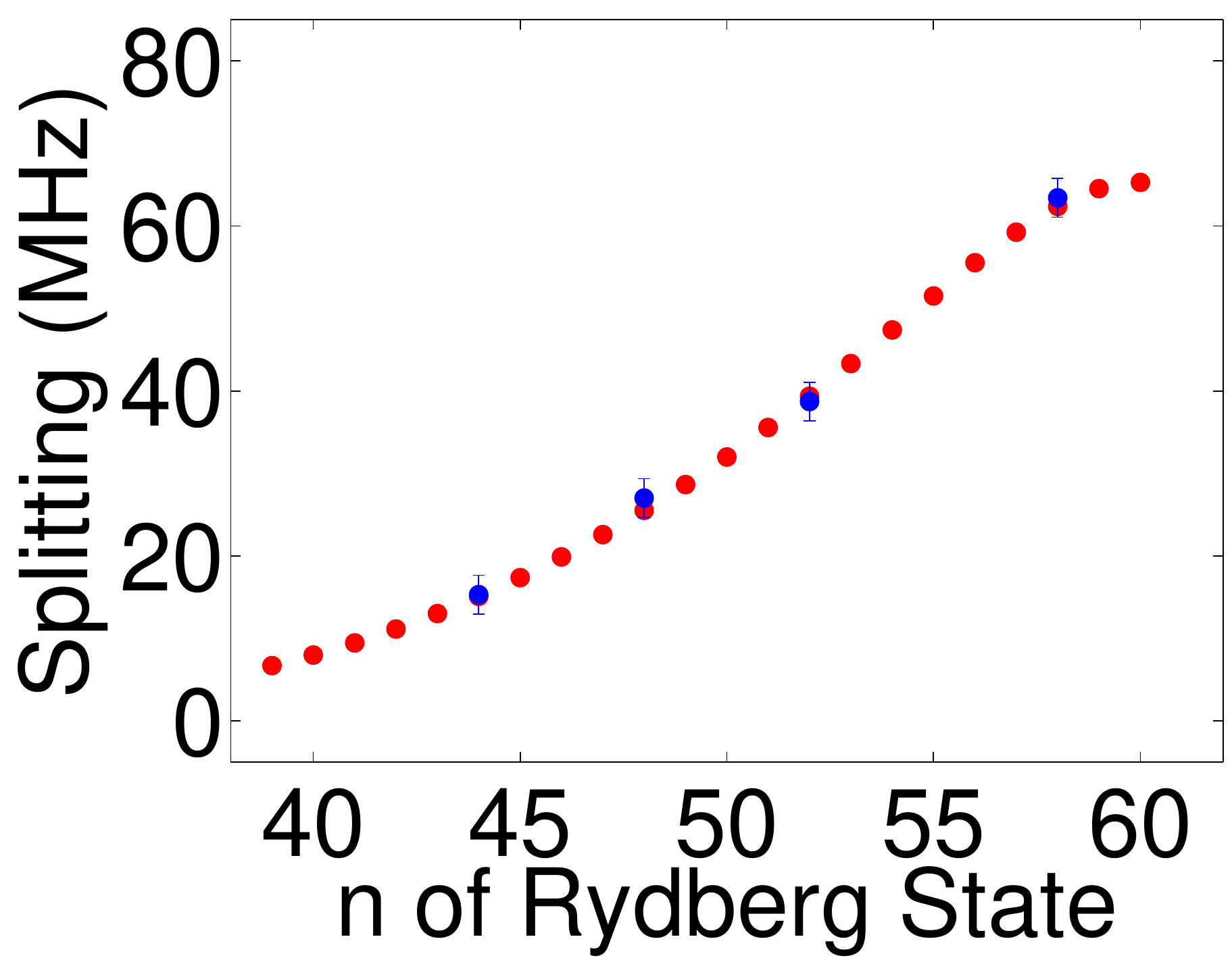}}}
\caption{\label{fig:cryo_temp_map}(color online). Rydberg spectroscopy of the 48$\textit{D}_{5/2} $ $ |m_j| = 3/2,1/2$ states below the YBCO square obtained at $83\; \mathrm{K}$. The $|m_j| = 5/2$ is shifted out of the scanning region. The horizontal axis is the detuning of the coupling laser from the $5P_{3/2}$ to 58$\textit{D}_{5/2}$ transition. The vertical axis is the distance to the substrate. Fitting Eq.(\ref{equa_dipole}) to the data (black  lines) gives a surface dipole density of $-1.7 \times 10^{-11} \; \text{C m$^{-1}$} $ distributed on YSZ over a size of $13.3 \; \mathrm{mm}$ and $-2.3 \times 10^{-11} \; \text{C m$^{-1}$} $ on YBCO of size 1 mm. [Inset] Measured separation (blue) of the $|m_j|=3/2$ and $|m_j|=1/2$ versus different n-states at 2 mm distance from the chip surface is in good agreement with the theory for the Stark shift (red), showing that the splitting is caused solely by electric fields.}
\end{figure}

This electric field behaviour versus distance from the chip changes drastically when the chip is cooled to cryogenic temperatures. Fig. \ref{fig:cryo_temp_map} shows the ODR using the 48$\textit{D}_{5/2}$ state at a temperature of 83 K. Only two lines are visible in the figure, corresponding to the $|m_j|=3/2$ and $|m_j|=1/2$ states. The magnitude of the electric field shifts the $|m_j|=5/2$ manifold out of our accessible coupling beam detunings. While the field is still seen to increase close to the chip surface, there is a significant, relatively constant field at distances exceeding $2 \; \mathrm{mm}$. Such a change in electric field distribution is not expected to arise due to chemisorption, as it is already saturated at room temperatures due to the strength of the bond. The constant field at distances far from the chip suggests a much larger layer of adsorbates than the one present on the 1 mm YBCO square, indicating the presence of physisorption on both the YSZ and YBCO. The fit to the electric field distribution gives in addition to the dipoles due to chemisorption, a dipole density of $-1.7\times 10^{-11} \; \text{C m$^{-1}$}$ distributed on YSZ over a size of $13.3 \; \mathrm{mm}$ and $-2.3\times 10^{-11} \; \text{C m$^{-1}$} $ distributed on YBCO of size $1 \; \mathrm{mm}$. While the localized contribution is explained by the presence of the YBCO square, the larger distribution of dipoles is comparable to the size of our YSZ substrate ($20 \; \mathrm{mm}$). To rule out line splittings due to a magnetic field, we plot the separation of the $|m_j|=3/2$ and $|m_j|=1/2$ lines as a function of n-state (inset of Fig. \ref{fig:cryo_temp_map}). The obtained $n$ dependency is in good agreement with the theory for the  splitting between $|m_j|=3/2$ and $|m_j|=1/2$ of 1 V/cm, indicating that our observations result entirely from coupling to the electric field.

\begin{figure}[t]
  \centering
\includegraphics[width=.48\textwidth]{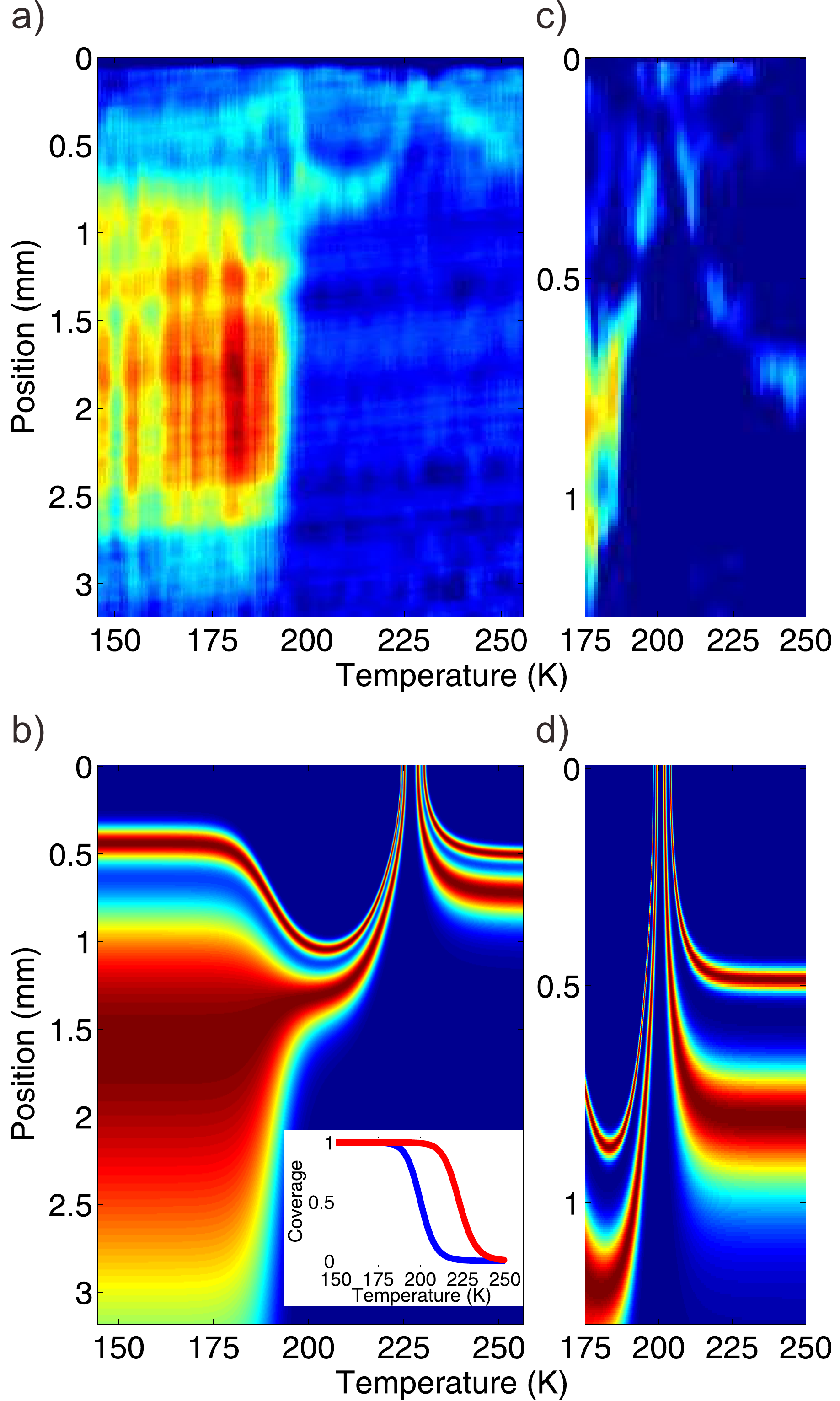}
\caption {(color online). a) Rydberg spectroscopy of the 52$\textit{D}_{5/2} $ $ |m_j| = 1/2$ states below the YBCO square in dependence of the temperature. The coupling beam detuning is 27 MHz while varying the temperature of the substrate. b) Simulation with a coupling beam detuning of 27 MHz and a linewidth of 7 MHz. Shown in the inset is the coverage dependence on the temperature for physisorbed atoms on YSZ (blue) and on YBCO (red) based on Eq.(\ref{equa_isobar}). The adsorption energies of physisorbed atoms on YSZ and YBCO are taken to be 790 meV and 880 meV respectively. c) Temperature dependent Rydberg spectroscopy of the 52$\textit{D}_{5/2} $ $ |m_j| = 1/2$ states with a chosen coupling beam detuning of 21 MHz. d) Simulation with adsorption energies of 703 meV and 783 meV on YSZ and on YBCO for coupling beam detuning of 21 MHz.}
\label{fig:eit_temp_dep}
\end{figure}

To study the temperature dependence of physisorption we look at the ODR of the 52$\textit{D}_{5/2}$ state with a fixed coupling beam detuning of $27 \; \mathrm{MHz}$. Motivated by the approach by Ranke et. al. \cite{ranke_phys}, we use the model developed by Langmuir to understand the dynamics of adsorbates with respect to temperature. With the adsorption of the gas phase and the desorption of the adsorbate phase in equilibrium, for constant pressure, the adsorbate coverage is given by \cite{book_langmuir_isobar}:
\begin{equation}
\frac{\theta_{ad}}{1-\theta_{ad}}=e^{\frac{\Delta E_{ad}}{k_BT_{sub}}}\frac{\hbar^3}{{k_B T_{gas}}(2\pi m k_B T_{gas})^{3/2}} P,
\label{equa_isobar}
\end{equation}
where $\theta_{ad}$ is the coverage of the adsorbates, $T_{sub}$ is the temperature of the substrate, $T_{gas}$ is the temperature of the gas phase, $P$ is the pressure and $\Delta E_{ad}$ is the energy of adsorption governing the temperature region where desorption starts to occur. Shown in Fig. \ref{fig:eit_temp_dep}a is the ODR versus distance from the chip surface as the chip is heated from cryogenic to room temperature. At 250 K, we only see a narrow region at which the $27 \; \mathrm{MHz}$ detuning matches the EIT resonance due to the strong spatial dependence of the electric field. This detuning translates to an electric field of 1.07 V/cm. As the chip temperature increases from 200 K, this resonance moves closer to the chip surface indicating a weakening of the adsorbate electric field. At approximately 230 K, the resonance vanishes implying the cancellation of the electric field. Above 230 K, the resonance reverts to the room temperature value. In Fig. \ref{fig:eit_temp_dep}c, we start with a non-saturated layer of physisorbed atoms at cryogenic temperature. To shift the resonance signal further from the diffraction lines close to the chip surface, we change the coupling beam detuning to 21 MHz. With a non-saturated physisorbed layer, the resultant electric field is smaller and the cancellation of the electric field occurs at a lower temperature at 202 K, showing the coverage dependence of adsorption energy \cite{book_langmuir_isobar}. 

The coverage dependent zero-crossing of the resonance and with this the electric field shows a change in the direction of the electric field. This process is reproduced by our model Eq.(\ref{equa_isobar}) in Fig. \ref{fig:eit_temp_dep}b and \ref{fig:eit_temp_dep}d for the temperature dependent adsorption process. As such, our data indicates that the dipole moment of the physisorbed rubidium atoms carries the opposite sign from those that are chemisorbed. Indeed, due to the difference in work function of the substrate and the ionization energy of rubidium atom, electrons are pulled from rubidium adatoms toward the substrate during chemisorption. In the case of physisorption on the other hand, electronic Pauli blocking will generally push electrons away from the substrate, leading to an opposite dipole moment compared to a chemisorbed adatom \cite{dipole_phys,Kamins}.

To explain the temperature dependence shown in Fig. \ref{fig:eit_temp_dep}a, we employ Eq.(\ref{equa_isobar}) giving the physisorbed coverage with respect to temperature (Fig. \ref{fig:eit_temp_dep}b). The onset of the large, constant electric field at 200 K is caused by physisorption on the large YSZ substrate. The transition at 230 K is a result of the onset of physisorption on the YBCO square. Fig. \ref{fig:eit_temp_dep}b shows the expected ODR behaviour taking these effects into account, with a coupling beam linewidth of $7 \; \mathrm{MHz}$. The theory is in close agreement with the experimental data, with the exception of the lines closer to the surface, as they are too weak to be detected in the experiment. The two distinct onsets of physisorption of YSZ and YBCO arise from a difference in adsorption energies for these materials. With our vacuum pressure of $P=4\times 10^{-10}$ mbar and a temperature of $T_{gas}=293$ K the adsorption energies of $^{87}$Rb are fitted to be $790 \; \mathrm{meV}$ and $880 \; \mathrm{meV}$ for YSZ and YBCO respectively. The simulated temperature dependence of the EIT resonance shown in Fig. \ref{fig:eit_temp_dep}d is obtained with the fitted adsorption energies of $703 \; \mathrm{meV}$ and $783 \; \mathrm{meV}$.

In conclusion, we have observed a strong change in the adsorbate density on a YBCO/YSZ atom chip at cryogenic temperatures due to physisorption. The weak chemisorption of rubidium atoms on YSZ observed suggests that, through the appropriate choice of substrate, the van der Waals interaction is the dominant factor governing the behaviour of adsorption of atoms onto the chip. The magnitude of the electric field resultant from this unavoidable physisorption is high enough to hinder controlled electric coupling between a superconducting solid state device and an atomic quantum system. However, our studies show that the physisorbed rubidium atom has a dipole moment of opposite sign compared to a chemisorbed rubidium atom, allowing for the possibility of cancelling out the electric field from chemisorbed atoms. In addition, the physisorbed atoms may serve as a potential candidate for studies of atom surface interactions at close distances.

This work was funded by the Centre for Quantum Technologies, Singapore. We would like to thank J. Fort\'{a}gh and H. Hattermann for helpful discussions.


\begin{thebibliography}{28}%
\makeatletter
\providecommand \@ifxundefined [1]{%
 \@ifx{#1\undefined}
}%
\providecommand \@ifnum [1]{%
 \ifnum #1\expandafter \@firstoftwo
 \else \expandafter \@secondoftwo
 \fi
}%
\providecommand \@ifx [1]{%
 \ifx #1\expandafter \@firstoftwo
 \else \expandafter \@secondoftwo
 \fi
}%
\providecommand \natexlab [1]{#1}%
\providecommand \enquote  [1]{``#1''}%
\providecommand \bibnamefont  [1]{#1}%
\providecommand \bibfnamefont [1]{#1}%
\providecommand \citenamefont [1]{#1}%
\providecommand \href@noop [0]{\@secondoftwo}%
\providecommand \href [0]{\begingroup \@sanitize@url \@href}%
\providecommand \@href[1]{\@@startlink{#1}\@@href}%
\providecommand \@@href[1]{\endgroup#1\@@endlink}%
\providecommand \@sanitize@url [0]{\catcode `\\12\catcode `\$12\catcode
  `\&12\catcode `\#12\catcode `\^12\catcode `\_12\catcode `\%12\relax}%
\providecommand \@@startlink[1]{}%
\providecommand \@@endlink[0]{}%
\providecommand \url  [0]{\begingroup\@sanitize@url \@url }%
\providecommand \@url [1]{\endgroup\@href {#1}{\urlprefix }}%
\providecommand \urlprefix  [0]{URL }%
\providecommand \Eprint [0]{\href }%
\providecommand \doibase [0]{http://dx.doi.org/}%
\providecommand \selectlanguage [0]{\@gobble}%
\providecommand \bibinfo  [0]{\@secondoftwo}%
\providecommand \bibfield  [0]{\@secondoftwo}%
\providecommand \translation [1]{[#1]}%
\providecommand \BibitemOpen [0]{}%
\providecommand \bibitemStop [0]{}%
\providecommand \bibitemNoStop [0]{.\EOS\space}%
\providecommand \EOS [0]{\spacefactor3000\relax}%
\providecommand \BibitemShut  [1]{\csname bibitem#1\endcsname}%
\let\auto@bib@innerbib\@empty
\bibitem [{\citenamefont {Fort\'agh}\ and\ \citenamefont
  {Zimmermann}(2007)}]{RevModPhys.79.235}%
  \BibitemOpen
  \bibfield  {author} {\bibinfo {author} {\bibfnamefont {J.}~\bibnamefont
  {Fort\'agh}}\ and\ \bibinfo {author} {\bibfnamefont {C.}~\bibnamefont
  {Zimmermann}},\ }\href {\doibase 10.1103/RevModPhys.79.235} {\bibfield
  {journal} {\bibinfo  {journal} {Rev. Mod. Phys.}\ }\textbf {\bibinfo {volume}
  {79}},\ \bibinfo {pages} {235} (\bibinfo {year} {2007})}\BibitemShut
  {NoStop}%
\bibitem [{\citenamefont {Reichel}\ and\ \citenamefont
  {Vuleti\'{c}}(2011)}]{AtomChips_Book}%
  \BibitemOpen
  \bibfield  {author} {\bibinfo {author} {\bibfnamefont {J.}~\bibnamefont
  {Reichel}}\ and\ \bibinfo {author} {\bibfnamefont {V.}~\bibnamefont
  {Vuleti\'{c}}},\ }\href@noop {} {\emph {\bibinfo {title} {Atom chips}}}\
  (\bibinfo  {publisher} {Wiley-VCH},\ \bibinfo {address} {Weinheim, Germany},\
  \bibinfo {year} {2011})\ pp.\ \bibinfo {pages} {xx, 425 p.}\BibitemShut
  {Stop}%
\bibitem [{\citenamefont {Treutlein}\ \emph {et~al.}(2004)\citenamefont
  {Treutlein}, \citenamefont {Hommelhoff}, \citenamefont {Steinmetz},
  \citenamefont {H\"ansch},\ and\ \citenamefont
  {Reichel}}]{PhysRevLett.92.203005}%
  \BibitemOpen
  \bibfield  {author} {\bibinfo {author} {\bibfnamefont {P.}~\bibnamefont
  {Treutlein}}, \bibinfo {author} {\bibfnamefont {P.}~\bibnamefont
  {Hommelhoff}}, \bibinfo {author} {\bibfnamefont {T.}~\bibnamefont
  {Steinmetz}}, \bibinfo {author} {\bibfnamefont {T.~W.}\ \bibnamefont
  {H\"ansch}}, \ and\ \bibinfo {author} {\bibfnamefont {J.}~\bibnamefont
  {Reichel}},\ }\href {\doibase 10.1103/PhysRevLett.92.203005} {\bibfield
  {journal} {\bibinfo  {journal} {Phys. Rev. Lett.}\ }\textbf {\bibinfo
  {volume} {92}},\ \bibinfo {pages} {203005} (\bibinfo {year}
  {2004})}\BibitemShut {NoStop}%
\bibitem [{\citenamefont {S\o{}rensen}\ \emph {et~al.}(2004)\citenamefont
  {S\o{}rensen}, \citenamefont {van~der Wal}, \citenamefont {Childress},\ and\
  \citenamefont {Lukin}}]{PhysRevLett.92.063601}%
  \BibitemOpen
  \bibfield  {author} {\bibinfo {author} {\bibfnamefont {A.~S.}\ \bibnamefont
  {S\o{}rensen}}, \bibinfo {author} {\bibfnamefont {C.~H.}\ \bibnamefont
  {van~der Wal}}, \bibinfo {author} {\bibfnamefont {L.~I.}\ \bibnamefont
  {Childress}}, \ and\ \bibinfo {author} {\bibfnamefont {M.~D.}\ \bibnamefont
  {Lukin}},\ }\href {\doibase 10.1103/PhysRevLett.92.063601} {\bibfield
  {journal} {\bibinfo  {journal} {Phys. Rev. Lett.}\ }\textbf {\bibinfo
  {volume} {92}},\ \bibinfo {pages} {063601} (\bibinfo {year}
  {2004})}\BibitemShut {NoStop}%
\bibitem [{\citenamefont {Andr\'{e}}\ \emph {et~al.}(2006)\citenamefont
  {Andr\'{e}}, \citenamefont {DeMille}, \citenamefont {Doyle}, \citenamefont
  {Lukin}, \citenamefont {Maxwell}, \citenamefont {Rabl}, \citenamefont
  {Schoelkopf},\ and\ \citenamefont {Zoller}}]{NaturePhysics.9.636}%
  \BibitemOpen
  \bibfield  {author} {\bibinfo {author} {\bibfnamefont {A.}~\bibnamefont
  {Andr\'{e}}}, \bibinfo {author} {\bibfnamefont {D.}~\bibnamefont {DeMille}},
  \bibinfo {author} {\bibfnamefont {J.~M.}\ \bibnamefont {Doyle}}, \bibinfo
  {author} {\bibfnamefont {M.~D.}\ \bibnamefont {Lukin}}, \bibinfo {author}
  {\bibfnamefont {S.~E.}\ \bibnamefont {Maxwell}}, \bibinfo {author}
  {\bibfnamefont {P.}~\bibnamefont {Rabl}}, \bibinfo {author} {\bibfnamefont
  {R.~J.}\ \bibnamefont {Schoelkopf}}, \ and\ \bibinfo {author} {\bibfnamefont
  {P.}~\bibnamefont {Zoller}},\ }\href {\doibase 10.1038/nphys386} {\bibfield
  {journal} {\bibinfo  {journal} {Nature Physics}\ }\textbf {\bibinfo {volume}
  {2}},\ \bibinfo {pages} {636} (\bibinfo {year} {2006})}\BibitemShut {NoStop}%
\bibitem [{\citenamefont {Jaksch}\ \emph {et~al.}(2000)\citenamefont {Jaksch},
  \citenamefont {Cirac}, \citenamefont {Zoller}, \citenamefont {Rolston},
  \citenamefont {C\^ot\'e},\ and\ \citenamefont {Lukin}}]{PhysRevLett.85.2208}%
  \BibitemOpen
  \bibfield  {author} {\bibinfo {author} {\bibfnamefont {D.}~\bibnamefont
  {Jaksch}}, \bibinfo {author} {\bibfnamefont {J.~I.}\ \bibnamefont {Cirac}},
  \bibinfo {author} {\bibfnamefont {P.}~\bibnamefont {Zoller}}, \bibinfo
  {author} {\bibfnamefont {S.~L.}\ \bibnamefont {Rolston}}, \bibinfo {author}
  {\bibfnamefont {R.}~\bibnamefont {C\^ot\'e}}, \ and\ \bibinfo {author}
  {\bibfnamefont {M.~D.}\ \bibnamefont {Lukin}},\ }\href {\doibase
  10.1103/PhysRevLett.85.2208} {\bibfield  {journal} {\bibinfo  {journal}
  {Phys. Rev. Lett.}\ }\textbf {\bibinfo {volume} {85}},\ \bibinfo {pages}
  {2208} (\bibinfo {year} {2000})}\BibitemShut {NoStop}%
\bibitem [{\citenamefont {Lukin}\ \emph {et~al.}(2001)\citenamefont {Lukin},
  \citenamefont {Fleischhauer}, \citenamefont {Cote}, \citenamefont {Duan},
  \citenamefont {Jaksch}, \citenamefont {Cirac},\ and\ \citenamefont
  {Zoller}}]{PhysRevLett.87.037901}%
  \BibitemOpen
  \bibfield  {author} {\bibinfo {author} {\bibfnamefont {M.~D.}\ \bibnamefont
  {Lukin}}, \bibinfo {author} {\bibfnamefont {M.}~\bibnamefont {Fleischhauer}},
  \bibinfo {author} {\bibfnamefont {R.}~\bibnamefont {Cote}}, \bibinfo {author}
  {\bibfnamefont {L.~M.}\ \bibnamefont {Duan}}, \bibinfo {author}
  {\bibfnamefont {D.}~\bibnamefont {Jaksch}}, \bibinfo {author} {\bibfnamefont
  {J.~I.}\ \bibnamefont {Cirac}}, \ and\ \bibinfo {author} {\bibfnamefont
  {P.}~\bibnamefont {Zoller}},\ }\href {\doibase 10.1103/PhysRevLett.87.037901}
  {\bibfield  {journal} {\bibinfo  {journal} {Phys. Rev. Lett.}\ }\textbf
  {\bibinfo {volume} {87}},\ \bibinfo {pages} {037901} (\bibinfo {year}
  {2001})}\BibitemShut {NoStop}%
\bibitem [{\citenamefont {Saffman}\ \emph {et~al.}(2010)\citenamefont
  {Saffman}, \citenamefont {Walker},\ and\ \citenamefont
  {M\o{}lmer}}]{RevModPhys.82.2313}%
  \BibitemOpen
  \bibfield  {author} {\bibinfo {author} {\bibfnamefont {M.}~\bibnamefont
  {Saffman}}, \bibinfo {author} {\bibfnamefont {T.~G.}\ \bibnamefont {Walker}},
  \ and\ \bibinfo {author} {\bibfnamefont {K.}~\bibnamefont {M\o{}lmer}},\
  }\href {\doibase 10.1103/RevModPhys.82.2313} {\bibfield  {journal} {\bibinfo
  {journal} {Rev. Mod. Phys.}\ }\textbf {\bibinfo {volume} {82}},\ \bibinfo
  {pages} {2313} (\bibinfo {year} {2010})}\BibitemShut {NoStop}%
\bibitem [{\citenamefont {Petrosyan}\ and\ \citenamefont
  {Fleischhauer}(2008)}]{PhysRevLett.100.170501}%
  \BibitemOpen
  \bibfield  {author} {\bibinfo {author} {\bibfnamefont {D.}~\bibnamefont
  {Petrosyan}}\ and\ \bibinfo {author} {\bibfnamefont {M.}~\bibnamefont
  {Fleischhauer}},\ }\href {\doibase 10.1103/PhysRevLett.100.170501} {\bibfield
   {journal} {\bibinfo  {journal} {Phys. Rev. Lett.}\ }\textbf {\bibinfo
  {volume} {100}},\ \bibinfo {pages} {170501} (\bibinfo {year}
  {2008})}\BibitemShut {NoStop}%
\bibitem [{\citenamefont {Petrosyan}\ \emph {et~al.}(2009)\citenamefont
  {Petrosyan}, \citenamefont {Bensky}, \citenamefont {Kurizki}, \citenamefont
  {Mazets}, \citenamefont {Majer},\ and\ \citenamefont
  {Schmiedmayer}}]{PhysRevA.79.040304}%
  \BibitemOpen
  \bibfield  {author} {\bibinfo {author} {\bibfnamefont {D.}~\bibnamefont
  {Petrosyan}}, \bibinfo {author} {\bibfnamefont {G.}~\bibnamefont {Bensky}},
  \bibinfo {author} {\bibfnamefont {G.}~\bibnamefont {Kurizki}}, \bibinfo
  {author} {\bibfnamefont {I.}~\bibnamefont {Mazets}}, \bibinfo {author}
  {\bibfnamefont {J.}~\bibnamefont {Majer}}, \ and\ \bibinfo {author}
  {\bibfnamefont {J.}~\bibnamefont {Schmiedmayer}},\ }\href {\doibase
  10.1103/PhysRevA.79.040304} {\bibfield  {journal} {\bibinfo  {journal} {Phys.
  Rev. A}\ }\textbf {\bibinfo {volume} {79}},\ \bibinfo {pages} {040304}
  (\bibinfo {year} {2009})}\BibitemShut {NoStop}%
\bibitem [{\citenamefont {Hogan}\ \emph {et~al.}(2012)\citenamefont {Hogan},
  \citenamefont {Agner}, \citenamefont {Merkt}, \citenamefont {Thiele},
  \citenamefont {Filipp},\ and\ \citenamefont
  {Wallraff}}]{PhysRevLett.108.063004}%
  \BibitemOpen
  \bibfield  {author} {\bibinfo {author} {\bibfnamefont {S.~D.}\ \bibnamefont
  {Hogan}}, \bibinfo {author} {\bibfnamefont {J.~A.}\ \bibnamefont {Agner}},
  \bibinfo {author} {\bibfnamefont {F.}~\bibnamefont {Merkt}}, \bibinfo
  {author} {\bibfnamefont {T.}~\bibnamefont {Thiele}}, \bibinfo {author}
  {\bibfnamefont {S.}~\bibnamefont {Filipp}}, \ and\ \bibinfo {author}
  {\bibfnamefont {A.}~\bibnamefont {Wallraff}},\ }\href {\doibase
  10.1103/PhysRevLett.108.063004} {\bibfield  {journal} {\bibinfo  {journal}
  {Phys. Rev. Lett.}\ }\textbf {\bibinfo {volume} {108}},\ \bibinfo {pages}
  {063004} (\bibinfo {year} {2012})}\BibitemShut {NoStop}%
\bibitem [{\citenamefont {McGuirk}\ \emph {et~al.}(2004)\citenamefont
  {McGuirk}, \citenamefont {Harber}, \citenamefont {Obrecht},\ and\
  \citenamefont {Cornell}}]{cornell_2004}%
  \BibitemOpen
  \bibfield  {author} {\bibinfo {author} {\bibfnamefont {J.~M.}\ \bibnamefont
  {McGuirk}}, \bibinfo {author} {\bibfnamefont {D.~M.}\ \bibnamefont {Harber}},
  \bibinfo {author} {\bibfnamefont {J.~M.}\ \bibnamefont {Obrecht}}, \ and\
  \bibinfo {author} {\bibfnamefont {E.~A.}\ \bibnamefont {Cornell}},\ }\href
  {\doibase 10.1103/PhysRevA.69.062905} {\bibfield  {journal} {\bibinfo
  {journal} {Phys. Rev. A}\ }\textbf {\bibinfo {volume} {69}},\ \bibinfo
  {pages} {062905} (\bibinfo {year} {2004})}\BibitemShut {NoStop}%
\bibitem [{\citenamefont {Obrecht}\ \emph {et~al.}(2007)\citenamefont
  {Obrecht}, \citenamefont {Wild},\ and\ \citenamefont
  {Cornell}}]{cornell_2007}%
  \BibitemOpen
  \bibfield  {author} {\bibinfo {author} {\bibfnamefont {J.~M.}\ \bibnamefont
  {Obrecht}}, \bibinfo {author} {\bibfnamefont {R.~J.}\ \bibnamefont {Wild}}, \
  and\ \bibinfo {author} {\bibfnamefont {E.~A.}\ \bibnamefont {Cornell}},\
  }\href {\doibase 10.1103/PhysRevA.75.062903} {\bibfield  {journal} {\bibinfo
  {journal} {Phys. Rev. A}\ }\textbf {\bibinfo {volume} {75}},\ \bibinfo
  {pages} {062903} (\bibinfo {year} {2007})}\BibitemShut {NoStop}%
\bibitem [{\citenamefont {Tauschinsky}\ \emph {et~al.}(2010)\citenamefont
  {Tauschinsky}, \citenamefont {Thijssen}, \citenamefont {Whitlock},
  \citenamefont {van Linden van~den Heuvell},\ and\ \citenamefont
  {Spreeuw}}]{spreeuw_2010}%
  \BibitemOpen
  \bibfield  {author} {\bibinfo {author} {\bibfnamefont {A.}~\bibnamefont
  {Tauschinsky}}, \bibinfo {author} {\bibfnamefont {R.~M.~T.}\ \bibnamefont
  {Thijssen}}, \bibinfo {author} {\bibfnamefont {S.}~\bibnamefont {Whitlock}},
  \bibinfo {author} {\bibfnamefont {H.~B.}\ \bibnamefont {van Linden van~den
  Heuvell}}, \ and\ \bibinfo {author} {\bibfnamefont {R.~J.~C.}\ \bibnamefont
  {Spreeuw}},\ }\href {\doibase 10.1103/PhysRevA.81.063411} {\bibfield
  {journal} {\bibinfo  {journal} {Phys. Rev. A}\ }\textbf {\bibinfo {volume}
  {81}},\ \bibinfo {pages} {063411} (\bibinfo {year} {2010})}\BibitemShut
  {NoStop}%
\bibitem [{\citenamefont {Abel}\ \emph {et~al.}(2011)\citenamefont {Abel},
  \citenamefont {Carr}, \citenamefont {Krohn},\ and\ \citenamefont
  {Adams}}]{cs_adams_ads}%
  \BibitemOpen
  \bibfield  {author} {\bibinfo {author} {\bibfnamefont {R.~P.}\ \bibnamefont
  {Abel}}, \bibinfo {author} {\bibfnamefont {C.}~\bibnamefont {Carr}}, \bibinfo
  {author} {\bibfnamefont {U.}~\bibnamefont {Krohn}}, \ and\ \bibinfo {author}
  {\bibfnamefont {C.~S.}\ \bibnamefont {Adams}},\ }\href {\doibase
  10.1103/PhysRevA.84.023408} {\bibfield  {journal} {\bibinfo  {journal} {Phys.
  Rev. A}\ }\textbf {\bibinfo {volume} {84}},\ \bibinfo {pages} {023408}
  (\bibinfo {year} {2011})}\BibitemShut {NoStop}%
\bibitem [{\citenamefont {Hattermann}\ \emph {et~al.}(2012)\citenamefont
  {Hattermann}, \citenamefont {Mack}, \citenamefont {Karlewski}, \citenamefont
  {Jessen}, \citenamefont {Cano},\ and\ \citenamefont
  {Fort\'agh}}]{fortagh_2012}%
  \BibitemOpen
  \bibfield  {author} {\bibinfo {author} {\bibfnamefont {H.}~\bibnamefont
  {Hattermann}}, \bibinfo {author} {\bibfnamefont {M.}~\bibnamefont {Mack}},
  \bibinfo {author} {\bibfnamefont {F.}~\bibnamefont {Karlewski}}, \bibinfo
  {author} {\bibfnamefont {F.}~\bibnamefont {Jessen}}, \bibinfo {author}
  {\bibfnamefont {D.}~\bibnamefont {Cano}}, \ and\ \bibinfo {author}
  {\bibfnamefont {J.}~\bibnamefont {Fort\'agh}},\ }\href {\doibase
  10.1103/PhysRevA.86.022511} {\bibfield  {journal} {\bibinfo  {journal} {Phys.
  Rev. A}\ }\textbf {\bibinfo {volume} {86}},\ \bibinfo {pages} {022511}
  (\bibinfo {year} {2012})}\BibitemShut {NoStop}%
\bibitem [{\citenamefont {Sattler}(2010)}]{book_nanophysics_clusters}%
  \BibitemOpen
  \bibfield  {author} {\bibinfo {author} {\bibfnamefont {K.}~\bibnamefont
  {Sattler}},\ }\enquote {\bibinfo {title} {Cluster-substrate interaction},}\
  in\ \href@noop {} {\emph {\bibinfo {booktitle} {Handbook of Nanophysics:
  Clusters and Fullerenes}}}\ (\bibinfo  {publisher} {Taylor \& Francis},\
  \bibinfo {year} {2010})\ Chap.~\bibinfo {chapter} {18}\BibitemShut {NoStop}%
\bibitem [{\citenamefont {Ibach}(2006)}]{book_langmuir_isobar}%
  \BibitemOpen
  \bibfield  {author} {\bibinfo {author} {\bibfnamefont {H.}~\bibnamefont
  {Ibach}},\ }\enquote {\bibinfo {title} {Adsorption},}\ in\ \href@noop {}
  {\emph {\bibinfo {booktitle} {Physics of Surfaces and Interfaces}}}\
  (\bibinfo  {publisher} {Springer},\ \bibinfo {year} {2006})\ Chap.~\bibinfo
  {chapter} {6}\BibitemShut {NoStop}%
\bibitem [{\citenamefont {Fleischhauer}\ \emph {et~al.}(2005)\citenamefont
  {Fleischhauer}, \citenamefont {Imamoglu},\ and\ \citenamefont
  {Marangos}}]{RevModPhys.77.633}%
  \BibitemOpen
  \bibfield  {author} {\bibinfo {author} {\bibfnamefont {M.}~\bibnamefont
  {Fleischhauer}}, \bibinfo {author} {\bibfnamefont {A.}~\bibnamefont
  {Imamoglu}}, \ and\ \bibinfo {author} {\bibfnamefont {J.~P.}\ \bibnamefont
  {Marangos}},\ }\href {\doibase 10.1103/RevModPhys.77.633} {\bibfield
  {journal} {\bibinfo  {journal} {Rev. Mod. Phys.}\ }\textbf {\bibinfo {volume}
  {77}},\ \bibinfo {pages} {633} (\bibinfo {year} {2005})}\BibitemShut
  {NoStop}%
\bibitem [{\citenamefont {M\"{u}ller}\ \emph {et~al.}(2010)\citenamefont
  {M\"{u}ller}, \citenamefont {Zhang}, \citenamefont {Fermani}, \citenamefont
  {Chan}, \citenamefont {Wang}, \citenamefont {Zhang}, \citenamefont {Lim},\
  and\ \citenamefont {Dumke}}]{atoms-vortices}%
  \BibitemOpen
  \bibfield  {author} {\bibinfo {author} {\bibfnamefont {T.}~\bibnamefont
  {M\"{u}ller}}, \bibinfo {author} {\bibfnamefont {B.}~\bibnamefont {Zhang}},
  \bibinfo {author} {\bibfnamefont {R.}~\bibnamefont {Fermani}}, \bibinfo
  {author} {\bibfnamefont {K.~S.}\ \bibnamefont {Chan}}, \bibinfo {author}
  {\bibfnamefont {Z.~W.}\ \bibnamefont {Wang}}, \bibinfo {author}
  {\bibfnamefont {C.~B.}\ \bibnamefont {Zhang}}, \bibinfo {author}
  {\bibfnamefont {M.~J.}\ \bibnamefont {Lim}}, \ and\ \bibinfo {author}
  {\bibfnamefont {R.}~\bibnamefont {Dumke}},\ }\href
  {http://stacks.iop.org/1367-2630/12/i=4/a=043016} {\bibfield  {journal}
  {\bibinfo  {journal} {New Journal of Physics}\ }\textbf {\bibinfo {volume}
  {12}},\ \bibinfo {pages} {043016} (\bibinfo {year} {2010})}\BibitemShut
  {NoStop}%
\bibitem [{\citenamefont {M\"uller}\ \emph {et~al.}(2010)\citenamefont
  {M\"uller}, \citenamefont {Zhang}, \citenamefont {Fermani}, \citenamefont
  {Chan}, \citenamefont {Lim},\ and\ \citenamefont
  {Dumke}}]{programmable_vortices}%
  \BibitemOpen
  \bibfield  {author} {\bibinfo {author} {\bibfnamefont {T.}~\bibnamefont
  {M\"uller}}, \bibinfo {author} {\bibfnamefont {B.}~\bibnamefont {Zhang}},
  \bibinfo {author} {\bibfnamefont {R.}~\bibnamefont {Fermani}}, \bibinfo
  {author} {\bibfnamefont {K.~S.}\ \bibnamefont {Chan}}, \bibinfo {author}
  {\bibfnamefont {M.~J.}\ \bibnamefont {Lim}}, \ and\ \bibinfo {author}
  {\bibfnamefont {R.}~\bibnamefont {Dumke}},\ }\href {\doibase
  10.1103/PhysRevA.81.053624} {\bibfield  {journal} {\bibinfo  {journal} {Phys.
  Rev. A}\ }\textbf {\bibinfo {volume} {81}},\ \bibinfo {pages} {053624}
  (\bibinfo {year} {2010})}\BibitemShut {NoStop}%
\bibitem [{\citenamefont {Siercke}\ \emph {et~al.}(2012)\citenamefont
  {Siercke}, \citenamefont {Chan}, \citenamefont {Zhang}, \citenamefont
  {Beian}, \citenamefont {Lim},\ and\ \citenamefont
  {Dumke}}]{self-sufficient-exp}%
  \BibitemOpen
  \bibfield  {author} {\bibinfo {author} {\bibfnamefont {M.}~\bibnamefont
  {Siercke}}, \bibinfo {author} {\bibfnamefont {K.~S.}\ \bibnamefont {Chan}},
  \bibinfo {author} {\bibfnamefont {B.}~\bibnamefont {Zhang}}, \bibinfo
  {author} {\bibfnamefont {M.}~\bibnamefont {Beian}}, \bibinfo {author}
  {\bibfnamefont {M.~J.}\ \bibnamefont {Lim}}, \ and\ \bibinfo {author}
  {\bibfnamefont {R.}~\bibnamefont {Dumke}},\ }\href {\doibase
  10.1103/PhysRevA.85.041403} {\bibfield  {journal} {\bibinfo  {journal} {Phys.
  Rev. A}\ }\textbf {\bibinfo {volume} {85}},\ \bibinfo {pages} {041403}
  (\bibinfo {year} {2012})}\BibitemShut {NoStop}%
\bibitem [{\citenamefont {Zimmerman}\ \emph {et~al.}(1979)\citenamefont
  {Zimmerman}, \citenamefont {Littman}, \citenamefont {Kash},\ and\
  \citenamefont {Kleppner}}]{zimmerman_stark}%
  \BibitemOpen
  \bibfield  {author} {\bibinfo {author} {\bibfnamefont {M.~L.}\ \bibnamefont
  {Zimmerman}}, \bibinfo {author} {\bibfnamefont {M.~G.}\ \bibnamefont
  {Littman}}, \bibinfo {author} {\bibfnamefont {M.~M.}\ \bibnamefont {Kash}}, \
  and\ \bibinfo {author} {\bibfnamefont {D.}~\bibnamefont {Kleppner}},\ }\href
  {\doibase 10.1103/PhysRevA.20.2251} {\bibfield  {journal} {\bibinfo
  {journal} {Phys. Rev. A}\ }\textbf {\bibinfo {volume} {20}},\ \bibinfo
  {pages} {2251} (\bibinfo {year} {1979})}\BibitemShut {NoStop}%
\bibitem [{\citenamefont {Nonaka}\ \emph {et~al.}(2002)\citenamefont {Nonaka},
  \citenamefont {Shimizu}, \citenamefont {Arai}, \citenamefont {Kurokawa},\
  and\ \citenamefont {Ichimura}}]{ybco_work_func}%
  \BibitemOpen
  \bibfield  {author} {\bibinfo {author} {\bibfnamefont {H.}~\bibnamefont
  {Nonaka}}, \bibinfo {author} {\bibfnamefont {T.}~\bibnamefont {Shimizu}},
  \bibinfo {author} {\bibfnamefont {K.}~\bibnamefont {Arai}}, \bibinfo {author}
  {\bibfnamefont {A.}~\bibnamefont {Kurokawa}}, \ and\ \bibinfo {author}
  {\bibfnamefont {S.}~\bibnamefont {Ichimura}},\ }\href {\doibase
  10.1384/jsa.9.344} {\bibfield  {journal} {\bibinfo  {journal} {Journal of
  Surface Analysis}\ }\textbf {\bibinfo {volume} {9}},\ \bibinfo {pages} {344}
  (\bibinfo {year} {2002})}\BibitemShut {NoStop}%
\bibitem [{\citenamefont {Vohrer}\ \emph {et~al.}(1993)\citenamefont {Vohrer},
  \citenamefont {Wiemhöfer}, \citenamefont {Göpel}, \citenamefont {van
  Hassel},\ and\ \citenamefont {Burggraaf}}]{ysz_work_function}%
  \BibitemOpen
  \bibfield  {author} {\bibinfo {author} {\bibfnamefont {U.}~\bibnamefont
  {Vohrer}}, \bibinfo {author} {\bibfnamefont {H.-D.}\ \bibnamefont
  {Wiemhöfer}}, \bibinfo {author} {\bibfnamefont {W.}~\bibnamefont {Göpel}},
  \bibinfo {author} {\bibfnamefont {B.}~\bibnamefont {van Hassel}}, \ and\
  \bibinfo {author} {\bibfnamefont {A.}~\bibnamefont {Burggraaf}},\ }\href
  {\doibase 10.1016/0167-2738(93)90240-4} {\bibfield  {journal} {\bibinfo
  {journal} {Solid State Ionics}\ }\textbf {\bibinfo {volume} {59}},\ \bibinfo
  {pages} {141 } (\bibinfo {year} {1993})}\BibitemShut {NoStop}%
\bibitem [{\citenamefont {Ranke}\ and\ \citenamefont
  {Joseph}(2002)}]{ranke_phys}%
  \BibitemOpen
  \bibfield  {author} {\bibinfo {author} {\bibfnamefont {W.}~\bibnamefont
  {Ranke}}\ and\ \bibinfo {author} {\bibfnamefont {Y.}~\bibnamefont {Joseph}},\
  }\href {\doibase 10.1039/B200363E} {\bibfield  {journal} {\bibinfo  {journal}
  {Phys. Chem. Chem. Phys.}\ }\textbf {\bibinfo {volume} {4}},\ \bibinfo
  {pages} {2483} (\bibinfo {year} {2002})}\BibitemShut {NoStop}%
\bibitem [{\citenamefont {Bruch}\ and\ \citenamefont
  {Ruijgrok}(1979)}]{dipole_phys}%
  \BibitemOpen
  \bibfield  {author} {\bibinfo {author} {\bibfnamefont {L.~W.}\ \bibnamefont
  {Bruch}}\ and\ \bibinfo {author} {\bibfnamefont {T.~W.}\ \bibnamefont
  {Ruijgrok}},\ }\href {\doibase 10.1016/0039-6028(79)90304-2} {\bibfield
  {journal} {\bibinfo  {journal} {Surface Science}\ }\textbf {\bibinfo {volume}
  {79}},\ \bibinfo {pages} {509} (\bibinfo {year} {1979})}\BibitemShut
  {NoStop}%
\bibitem [{\citenamefont {Kamins}(1968)}]{Kamins}%
  \BibitemOpen
  \bibfield  {author} {\bibinfo {author} {\bibfnamefont {T.~I.}\ \bibnamefont
  {Kamins}},\ }\href {\doibase /10.1063/1.1655797} {\bibfield  {journal}
  {\bibinfo  {journal} {Journal of Applied Physics}\ }\textbf {\bibinfo
  {volume} {39}},\ \bibinfo {pages} {4529} (\bibinfo {year}
  {1968})}\BibitemShut {NoStop}%
\end{thebibliography}
\end{document}